\documentclass[runningheads]{llncs}

\usepackage{times}
\usepackage{soul}
\usepackage{url}
\usepackage[hidelinks]{hyperref}
\usepackage[utf8]{inputenc}
\usepackage[small]{caption}
\usepackage{graphicx}
\usepackage{mathrsfs} 
\usepackage{amsmath}
\usepackage{booktabs}
\usepackage{algorithm}
\usepackage{algorithmic}
\usepackage{adjustbox}
  \usepackage{todonotes}
\begin{document}
\title{Network-based methods for disease-gene prediction}
%
%
\author{Lorenzo Madeddu\inst{1, 4}\orcidID{0000-1111-2222-3333} \and
Giovanni Stilo\inst{2,4}\orcidID{0000-0002-2092-0213} \and
Paola Velardi\inst{3, 4}\orcidID{0000-0003-0884-1499}}
\authorrunning{L. Madeddu et al.}
%
\institute{
Translational and Precision Medicine Department, Sapienza Università di Roma, Roma, Italy 
\and
Department of Engineering and Information Sciences and Mathematics, Università degli Studi dell'Aquila, L'Aquila, Italy
\and
Computer Science Department, Sapienza Università di Roma, Roma, Italy
\and  
Intelligent Information Mining - Research Group\\
\url{http://iim.di.uniroma1.it}, \url{http://iim.disim.univaq.it}\\
\email{madeddu@di.uniroma1.it, giovanni.stilo@univaq.it, velardi@di.uniroma1.it}
}
\maketitle              
\begin{abstract}
We predict disease-genes relations on the Human Interactome network using a  methodology that jointly learns functional and  connectivity  patterns surrounding proteins. Contrary to other data structures, the Interactome is characterized by high incompleteness and absence of explicit negative knowledge, which makes predictive tasks particularly challenging. To exploit at best  latent information  in the network, we propose an extended version of random walks, named Random Watcher-Walker ($RW^2$),  which is able to learn rich representations of disease genes (or gene products) features.
Our method  successfully compares with the best known system for disease gene prediction,  and other state-of-the-art graph-based methods. 
We perform sensitivity analysis and apply  perturbations to ensure  robustness. In contrast with previous studies, our results demonstrate that connectivity alone is not sufficient to classify disease-related genes.
\end{abstract}
\section{Introduction}
\label{sec:intro}
  In the last decades, the evolution of medical knowledge has been supported by  academic research and  technological developments, on the one hand providing a continuously growing set of biomedical data and on the other hand revealing a complexity only perceived until now. In this context, biological networks have become a central hub of multidisciplinary research \cite{reviewbiological}, to address important challenges on both diagnostic and therapeutic aspects, such as  drug development and  disease classification \cite{drugrepurposing}\cite{drugrepositioning}\cite{diseaserelationships}.  

Network Medicine \cite{networkmedicine} (NM) is a relatively recent approach to analyze the complexity of biomolecular structures. NM proposes to exploit the network topology (e.g. the relations among biological entities)
and the network dynamics (e.g. the information flow across the network) to better understand  the pathogenic behavior of complex molecular interconnections, that  standard reductionist\footnote{according to reductionism, a single disease cause can be identified by decoupling the complex biological or medical phenomenons in multiple components.} approaches cannot detect.  A central finding of NM is the following \cite{networkmedicine}: "If a
gene or molecule is involved in a specific biochemical
process or disease, \textit{its direct interactors might also be
suspected to have some role in the same biochemical
process}. In line with this ‘local’ hypothesis,
proteins that are involved in the same disease show a
high propensity to interact with each other". Several studies have been published in support of this hypothesis, such as \cite{pub.1100212432}, \cite{Ghiassian2015}, \cite{Silberberg2017}  and others. 

It is important to stress the potential impact of network methods to progress in this field. In fact, traditional ways to assess the role of genes in diseases involve time-consuming and extremely expensive\footnote{https://www.genome.gov/27541954/dna-sequencing-costs-data/} statistical studies based on sequencing the DNA of a large number of patients affected by a given disease, known as Genome-Wide association studies (GWAS). In this context, \textit{network science and machine learning methods can be effective in reducing the number of alternatives to be explored} in clinical experiments. 

The objective of this paper is to contribute to the problem of predicting disease-related genes. We present  a graph-based approach, based on an extended notion of random walks, to extract  topological information and functional properties of local sub-structures within the human interactome network.  Detected patterns are then used to train a machine learning predictor. 
Our method  advances the state of the art, by successfully comparing with the best known system for disease gene prediction, and other known graph-based methods. Furthermore, our study brings new insight to the subject matter, showing that connectivity alone is not sufficient to reliably identify disease-related  genes. 

\section{Related Work}
\label{sec:related}
 Recent research fields such as System Biology and Network Medicine (NM) \cite{networkmedicine} has led to new approaches integrating the so called \textit{-omics} fields of study (genomics, proteomics and metabolomics) and network science. In these studies, complex physical and structural interactions between molecules are modeled as a graph structure, called \textit{Interactome}. 
As mentioned in the introduction, the driving idea of NM is that the study of network topology and  dynamics can accelerate discovery of new biological interactions and pathways \cite{chan}, which in turn will drive progress on disease treatments and personalized medicine. However,
the application of "standard" graph-based methods to the human Interactome is complicated by the fact that 
it is estimated that only 20\% of existing interactions are actually modeled \cite{Venkatesan}. Another problem is absence of explicit encoding of negative knowledge: when a relation is not present between two biological entities, we don't know if the relation actually does not exist or if it is still unknown\footnote{This is motivated by the inherent complexity of negative hypothesis testing in this field.}. 
Our intuition is that these specific characteristics of the Interactome network cause many commonly used graph-mining methods to be ineffective. For example, \cite{Aijun} and  \cite{Ghiassian2015} found that  community detection algorithms and centrality measures fail to identify relevant structures.
On the other hand, the interest of scientists in NM remains high, since a growing interdisciplinary effort  gives hope for an acceleration of results in this field.

Disease Gene Prediction (DGP) is a relevant, but still open, research topic, since the genetic bases of  diseases are largely unknown. Currently, only 10\% of genes have a known association with some disease \cite{networkmedicine}.  Genome-Wide associations studies (GWAS) have led to the collections of  such associations in databases, like OMIM \cite{omim} and DisGenNet \cite{disgenet}. However, as mentioned in the introduction, GWAS studies are very expensive and labor intensive.
Following the categorization in \cite{networkmedicine}, there exist several interactome-based approaches for the DGP problem,  divided into three categories: 
\begin{enumerate}
    \item Linkage Methods:
	Candidate disease genes (or gene \textit{products}\footnote{in gene prediction literature, proteins are commonly used in place of genes, since each gene tells the cell how to put together the building blocks for one specific protein. Therefore, studies in this field  consider protein-protein interaction networks in place of genes.}) are hypothesized to have a direct interaction with known disease genes in the interactome. For example, in \cite{oti}  candidate disease genes are identifyed in the direct neighborhood of known disease genes and then filtered  using their chromosomal locations (the authors check if the chromosomal location falls within one or more "disease loci").
	\item Module-based Methods:
	Candidate genes belonging to the same neighborhoods are more likely to be involved in the same diseases. These approaches are also know as "disease module-based", like for example
	\cite{Ghiassian2015,Silberberg2017}. Note that the notions of  "neighborhood" and "module" are vague here, and standard community detection algorithms do not work.
	Both approaches expand  disease modules by exploiting the structure of the network. The main idea of DIAMOnD \cite{Ghiassian2015} is based on the use of a \textit{connectivity significance} measure, designed to take advantage of the weak interconnection properties of the interactome. Using this metric, DIAMOnD first generates a connection ranking for each node, with respect to a chosen disease module. DIAMOnD works by iteratively  expanding a single disease module with the first ranked node identified in each iteration. Unlike DIAMOnD, Gladiator \cite{Silberberg2017} considers multiple disease modules simultaneously. Gladiator is based on the intuition that nodes with common phenotypes (common sets of symptoms) are also likely to share  molecular mechanisms\footnote{One of the problems of Gladiator is that the use of phenotipic information as a predictor of disease-gene relatedness for multi-factorial diseases is not assessed in literature.}. In order to predict gene-disease relationships, Gladiator uses a simulated annealing algorithm that considers both information on phenotypic similarity and protein interconnection. 
	\item  Diffusion Methods:
	are ranking approach based on the random-walk technique \cite{prince,randomwalkrestart,cipher,rwrh,hotnet2}. 
	For example, in order to find novel disease-gene candidates, \cite{randomwalkrestart} introduce random walk with restart, starting from genes known to be associated with a given disease. 
\end{enumerate}

In this paper, we introduce a fourth category, that we call Representation Learning Methods, consisting in the use of graph-based methods  \cite{node2vec}\cite{Kipf2016} to learn rich representations of disease genes (or gene products) \textit{features}, followed by a machine learning predictor for detecting candidate genes. 
As an example in this category, we present the
 Random Watcher-Walker ($RW^2$), described in Section \ref{sec:method}.

\section{Description of the Method}
\label{sec:method}
\begin{figure}[thpb]
\centering
\includegraphics[width=0.8\linewidth]{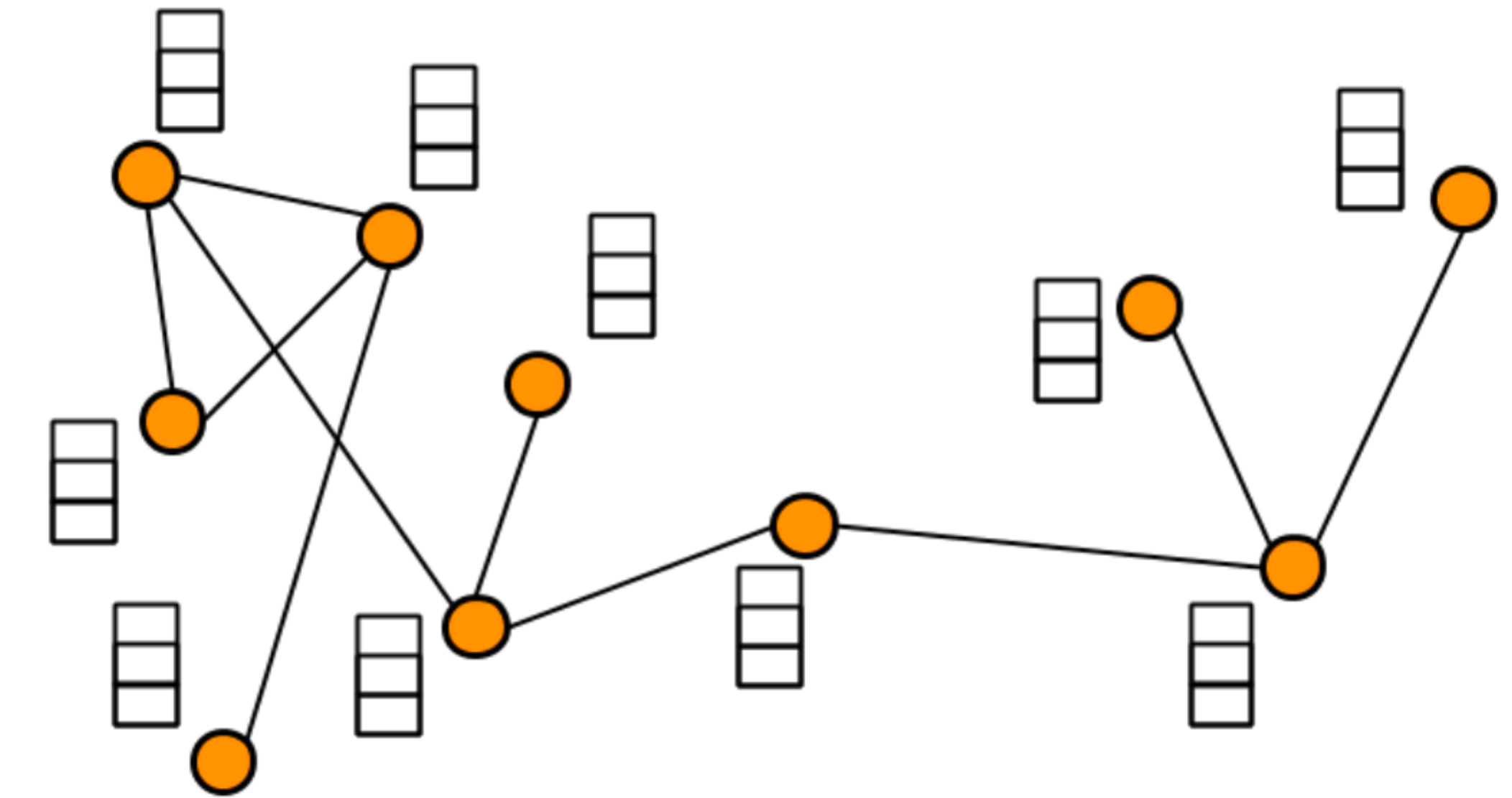}
\caption{ The network model}
\label{fig:graph}
\end{figure}

We predict disease genes using a graph-based methodology which jointly learns functional and  connectivity  patterns surrounding proteins in the human interactome. The network model G(V,E) is shown in Figure \ref{fig:graph}: nodes  $v \in V$ are proteins, and edges $e(i,j) \in E$, $i,j \in V$ are protein-protein interactions (as derived from three databases:  \textit{IntAct} \cite{intact}, \textit{BioGrid} \cite{biogrid}, \textit{MINT} \cite{mint}.  This structure is also referred to in literature as PPI (Protein-Protein Interaction) network\footnote{Note that different studies have used different knowledge bases to build the PPI.}. 

In our approach,  each node $v$ is further described by  a multi-dimensional feature vector $f(v)$. In the present study, besides \textit{diseases} (from OMIM \cite{omim}), we consider two additional relevant\footnote{following the indications of domain experts} dimensions:   \textit{tissues} (from the \textit{The Human Protein Atlas \cite{atlas}}) and \textit{pathways}\footnote{\url{https://en.wikipedia.org/wiki/Biological_pathway}} (from KEGG \cite{kegg}). For each dimension, a valued cell in $f(v)$  indicates that $v$
 is related  with a given disease, tissue or pathway category, hereafter denoted as \textit{"category label"} or \textit{label}. Category labels are provided in the same datasets (OMIM, etc.). 


The methodology to predict disease-related genes is in three steps: 

\noindent
\textbf{Step 1: Random Watcher Walker}: we collect \textit{network connectivity patterns} using a novel method, Random Watcher-Walker ($RW^2$). In $RW^2$, the walker, when landing on node $v$, "watches" the node features and selects one label at random with uniform probability in  those cells of $f(v)$ that are equals to 1. Next,  it "walks" with uniform probability to one of $v$'s neighbors. In this way, random walks embody both \textit{functional} features of traversed nodes (pathobiological properties such as related diseases, tissues and pathways), and \textit{structural} features (connected proteins in the interactome). 

$RW^2$ can be seen as a label sequence generation where, $v^e$ denote the $e^{th}$ node in the walk, and  $l^e$ denote the selected label of $v^e$. The generation process  satisfies the following distribution:

\[
P( v^e=x,\:l^e=a | v^{e-1}=y)=\\
\]
\[
\begin{cases}
{\pi(y,x)}\cdot{\sigma(x,a)} & \text{if }(y,x) \in E \text{ and $a$ is a label of x}\\
    0 & \text{otherwise}
\end{cases}
\]
\\
\noindent
where $\pi(y,x)$ is the normalized transition probability between nodes y and x; $\sigma_(x,a)$  is the normalized  probability of selecting the node-label $a$ in  $f(x)$.

\begin{figure}[thpb]
\centering
\includegraphics[width=0.95\linewidth]{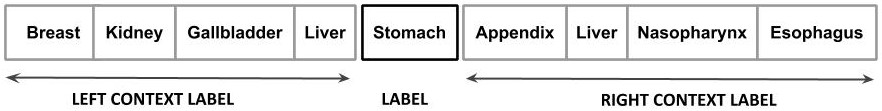}
\caption{ Example of context for the label "Stomach"}
\label{fig:context}
\end{figure}

Note that our Random Watcher Walker approach is meant to exploit one relevant findings of Network Medicine, the "modular" structure of diseases in the interactome, as summarized in Section \ref{sec:related}. Furthermore, our intuition is that random walks crossing nodes associated  with disease modules that are either close, or intersect each other in the interactome, should have similar label subsequences, since they are extracted from a similar neighborhood.  Given the "loose" notion of neighborhood implemented by random walks,  similarity patterns might be captured even in presence of highly incomplete knowledge.\\
\noindent
\textbf{Step 2: Label Embeddings}: collected network connectivity patterns are treated as "contexts" for individual labels, (as shown in Figure \ref{fig:context}) much in same way as sentences are contexts for individual words.
Context are used to train a Skip-Gram \cite{skipgram} model and learn \textit{label embeddings} ("dense" vector representations of each label).  Label embeddings are used to enrich the 
multidimensional feature vector $f(v)$
of each node of G: valued cells are replaced by the respective embedding vectors, producing the enriched feature matrix $\mathscr{F}(v)$.

\begin{figure}[thpb]
\centering
\includegraphics[width=.7\linewidth]{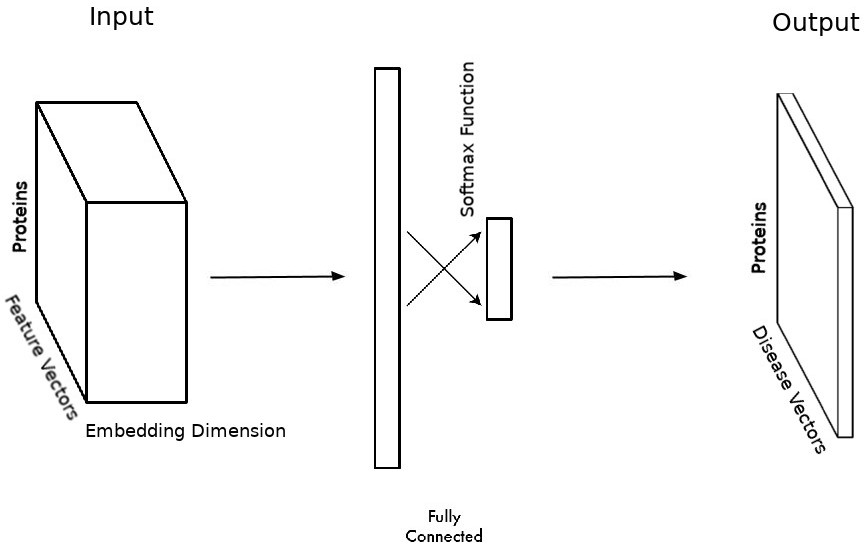}
\caption{Training the NN with feature matrices}
\label{fig:neural}
\end{figure}

\noindent
\textbf{Step 3: Training}: feature matrices $\mathscr{F}(v)$
are used to train a fully connected neural network (NN) with Softmax activation function, for predicting disease-gene associations (Figure \ref{fig:neural}).  The system's output is a ($|D|+1$)-dimensional \textit{probability vector}, where $|D|$ is the number of  considered disease labels and the additional class label is UNK, to state absence of known disease relations for a given node.\\ 


\section{Evaluation Methodology}
\label{sec:eval}
In this section we describe the dataset used for our experiments, the adopted data transformation methodology, and the experimental strategies and setup.\\

\noindent
\textbf{Dataset:} As we already remarked, biomedical data  suffer from two problems:
\begin{itemize}
    \item \textit{Absence of Negative Instances:}
    Biomedical data are not characterized by negative instances. Databases usually do not collect the information about the absence of a relationship between a gene and a disease.\\
   \item  \textit{Incompleteness:}
    The   high incompleteness of the human interactome  (see Section \ref{sec:intro}) implies that the information to be exploited can be very sparse.
    \end{itemize}
For this reason, evaluating on synthetic data 
may lead to inconclusive results, since the generated datasets can hardly satisfy the statistical properties of the real interactome. In our experiment, for the purpose of comparison, we use the same interactome (PPI) network used in  DIAMOnD \cite{Ghiassian2015}. In this network, there are 13458 proteins and 70 disease categories.  Only 1536 nodes are related to at least 1 disease, of which,  766 have co-morbidities (i.e., they are known to be related to more than one disease).  In addition to information on disease categories, we enrich feature vectors $f(v)$ with information on 57 tissue types\footnote{taken from \url{https://www.proteinatlas.org/humanproteome/tissue}} and 321 pathways. \\

\begin{figure}[thpb]
\centering
\includegraphics[width=.9\linewidth]{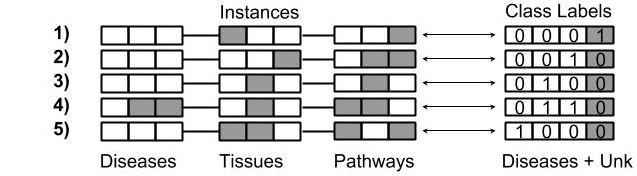}
\caption{Feature Matrices and ground-truth vectors}
\label{fig:before}
\end{figure}

\noindent
\textbf{Data Transformation:} Figure \ref{fig:before}  shows, for each node $v$ of the interactome, the enriched  multidimensional feature matrix $\mathscr{F}(v)$  (left) and the corresponding ground-truth output vectors $\mathscr{D}$ (right) to be used for training. Dark cells in $\mathscr{F}(v)$ represent embedding vectors associated to valued feature labels in the original $f(v)$, while white cells are zero vectors. The ($|D|+1$)-dimensional ground-truth  vector $\mathscr{D}$ has the $i^{th}$ cell equal to 1 if the node is known to be associated to the corresponding disease $d \in D$. The last cell of this vector is 1 if no diseases are known to be associated to the node.

\begin{figure}[thpb]
\centering
\includegraphics[width=.9\linewidth]{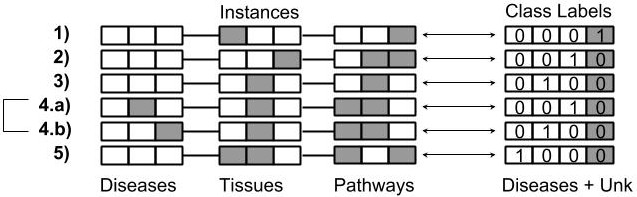}
\caption{Modified Feature Matrices  for the learning phase}
\label{fig:after}
\end{figure}
Clearly, the dataset in Figure \ref{fig:before} cannot be used for training, because the NN would trivially learn that if a disease vector is valued in $\mathscr{F}(v)$, then the corresponding cell of the output should be 1.
To avoid  \textit{trivial learning}, we train the  NN using modified feature matrices , as shown in Figure \ref{fig:after}. 
\begin{enumerate}
\item If a node $v$ is known to be related to a single disease $d$,  $D^v:\{d\}$, $|D^v|=1$ , then the corresponding 
embedding vector from the feature matrix  $\mathscr{F}(v)$ is replaced with a zero vector. For example, nodes 2), 3) and 5) of Figure  \ref{fig:before} are modified as in Figure \ref{fig:after}. Note that in this way a node with no valued cells in the \textit{disease} dimension (instances 1 and 2 of Figure \ref{fig:after}), can either be "unknown" - which corresponds to a $1$ in the last cell of the ground-truth vector - or known to be related with one disease. The NN  is therefore encouraged to learn distinguishing between these two cases based on the embedding vectors of the other dimensions, \textit{pathways} and \textit{tissues}; 

\item If a node $v$ is known to be related to $m$ diseases $D^v: \{d_1,\dots ,d_m\}$, $|D^v|=m$ , then its feature matrix  is duplicated into $m$  matrices. Each duplicated matrix  is associated to only one disease $d_k \in D^v$. In each duplicated feature matrix $\mathscr{F}(v)^k$ we replace the corresponding embedding vector of the associated disease $d_k$ with a zero vector, and we keep only the 1 associated with $d_k$ in the related ground-truth vector. 
For example, node 4 in Figure \ref{fig:before} is duplicated in 4.a and 4.b  in Figure \ref{fig:after}. This encourages the NN to learn also from \textit{co-morbidities}.
\end{enumerate}

\noindent
\textbf{Experimental Strategies:}
We first adopt a computationally expensive, but unbiased, strategy (\textit{method 1}). With method 1, the  Random Watcher-Walker $RW^2$ collects connectivity patterns from a "degraded" network from which 20\% of the information on node labels has been removed. Node labels are transformed as for in Figure \ref{fig:after}. 
Next, we learn label embeddings and train the NN network.  Note that in method 1, we repeat all the three steps of the algorithm (random walks, label embedding and NN training)  for  each of the $K$ different train/test split of the dataset used during the $K$-fold evaluation  of the system.
We also introduce a less computationally expensive strategy (\textit{method 2}) in which the previously described data transformation is performed \textit{after} label embedding without degrading the interactome at each fold evaluation. 
The dataset shown in Figure \ref{fig:after} is used to train the NN with 80\%  20\% split and 10-fold evaluation ($K=10$). However, learning the embeddings on the complete Interactome network  could potentially bias the results of the subsequent NN training phase.



\noindent
\textbf{Experimental setup:} Tables \ref{tab:neparameters1}  and \ref{tab:neparameters2} show the system parameters for our best experiment. Sensitivity to parameters is discussed in Section \ref{sub:robustness}.

\begin{table}[H]
\centering
\begin{tabular}{lc}
\toprule
Random Walk Parameter & Value\\
\midrule
Label Embedding Length & 128\\
Walk Length & 20\\
Number of random walks per node & 100\\
$p$ & 1\\
$q$ & 1\\
Skip-Gram context window & 3\\
Skip-Gram Epochs & 3\\
\bottomrule
\end{tabular}
\caption{Model parameters.}
\label{tab:neparameters1}
\end{table}
\vspace{-1.5em}
\begin{table}[H]
\centering
\begin{tabular}{lc}
\toprule
NN Parameter & Value\\
\midrule
Hidden Layer & 0\\
Activation Function & Softmax\\
Loss & Binary Cross-entropy\\
Optimizer & Adam\\
Batch Size & 100\\
Epochs & 5\\
\bottomrule
\end{tabular}
\caption{Model parameters.}
\label{tab:neparameters2}
\end{table}\vspace{-1.5em}
\noindent


\section{Experiments}\label{sec:exp}
\subsection{Comparison with other methods}
\label{sub:comparison}
Given the previously outlined characteristics  of biomedical data, evaluation measures such as \textit{precision, accuracy} and \textit{f-score} are ineffective. In line with other works \cite{Ghiassian2015}\cite{Silberberg2017} in this domain, we use  Recall@k, the fraction of correctly predicted items at rank k. In all our experiments, we set  the $k$ value of Recall@k to 1, since as explained in Section \ref{sec:intro}, the intended use of network methods in medicine is to exploit the results with the highest confidence, to narrow the scope of expensive and labor intensive clinical tests. 

We compare our system (trained according to methods 1 and 2, described in Section \ref{sec:eval}) with:
\begin{enumerate} 
\item A baseline method which uses only functional information, i.e. the feature vectors $f(v)$ without label embeddings. This corresponds to exploiting \textit{only functional (feature) similarity}.  
  
\item Node2Vec \cite{node2vec} combined with the NN Classifier. Node2Vec (N2V) is a state-of-the-art random walk algorithm for Node-Embedding generation. Contrary to our approach, Node2Vec uses a standard random-walk strategy which exploits only the connectivity patterns of a network. Consequently, Node2Vec generates embeddings containing \textit{only connectivity information}.
\item Graph Convolutional Networks (GCN) \cite{Kipf2016},  an inductive  semi-supervised method to classify nodes in a network, based on an efficient variant of convolutional neural networks, which operates directly on graphs.  GCN  handle both  the \textit{functional and  connectivity} information of the network but in a completely different way compared to our method.
We test GCN  with default parameters (as specified in \cite{Kipf2016}).
\item DIAMOnD, which is commonly considered the state of art and most cited study on disease gene prediction (see Section \ref{sec:related}). DIAMOnD, like N2V, exploits only connectivity information. 

\end{enumerate}

For all these methods, during the training phase we remove 20\% of the information concerning disease-node relationships, and use these data for testing. Each experiment is repeated 10 times with different splits of the learning and test set (10-fold cross evaluation).  Next, we compute the Recall@1  and   average over all folds.
Note that  computing Recall@1 for DIAMOnD is not straightforward. In DIAMOnD evaluation experiments, described in \cite{Ghiassian2015},  diseases are considered one at the time. For each disease $d$, they remove  node-disease associations from a given fraction of nodes $N'_d$ among those known to be related with $d$. Next, they apply an iterative method in which, at each iteration, they add a new node $n$ (the most likely node among those considered) to the current set of  nodes believed to be related to $d$. 
In their paper, the authors perform 200 iterations and lastly they compute the recall, i.e. the fraction of disease nodes retrieved by their method, among those ($N'_d$) that were initially removed. Although the authors do not explicitly set/report a k value for the Recall, we can assume that setting k=1 for their system is an $upper-bound$ of the real system performances. In our experiments, we use the software made available by the authors, and adopt exactly the same iterative methodology, removing 20\% of disease-node associations, like for the other compared methods.

The results of all comparative experiments are shown in Table \ref{tab:comparison}. 
The first two rows of Table \ref{tab:comparison} show the performance of our system.  The difference between methods 1 and 2 (see Section \ref{sec:eval}) is less than 1\% and is not statistically significant ($p=0.14$). This was expected, given the very sparse nature of the interactome graph. A comparison between the two methods was however necessary to exclude any learning bias. Given the negligible difference between the two methods, the less computationally intensive method 2 can be adopted instead of  method 1. 

Table \ref{tab:comparison} also shows that N2V and GCN perform very poorly: they are unable to identify patterns that help to predict disease nodes. As a result, they always predict the majority class (the label "unknown")
which  explains why the STD is zero.  
For DIAMOnD, we report the  recall value obtained as previously explained, averaging over all diseases, different splits, and 200 iterations for each experiment. 
We remark that in \cite{Ghiassian2015} the reported performance is higher, but limited to two diseases, lysosomal storage diseases and lipid metabolism disorders, that show higher density of the respective modules.  For the purpose of completeness, Table \ref{tab:papercomparison}  compares $RW^2$ and DIAMOnD on these very same diseases.  Furthermore, the authors of \cite{Silberberg2017}, in an experiment considering all diseases  (on a slightly different dataset), reported that DIAMOnD was "able to recover 13.3\% of the removed associations", which in line with the performance value (14\%)  in Table \ref{tab:comparison}.

\begin{table}[H]
\centering
\begin{tabular}{lcc}
\toprule
NN Parameter & R@1 & STD\\
\midrule
$RW^2$ (method 1) & 35\%  & 0.008\\
$RW^2$ (method 2) & 36\%  & 0.008\\
Baseline & 2\% & 0.002\\
N2V & 1\% & 0\\
GCN & 1\% & 0\\
DIAMOnD & 14\%* & 0.011\\
\bottomrule
\end{tabular}
\caption{Average Recall@1 and standard deviation over 10 folds. *Note that, as explained, for DIAMOnD the reported performances represent an upper-bound.}
\label{tab:comparison}

\end{table}
\vspace{-1.5em}
\begin{table}[H]
\centering
\begin{tabular}{lcc}
\toprule
Disease Module & $RW^2$ & DIAMOnD\\
\midrule
lysosomal storage diseases & 79\% & 50\%\\
lipid metabolism disorders & 32\% & 31\%\\
\bottomrule
\end{tabular}
\caption{Comparison on two diseases}
\label{tab:papercomparison}
\end{table}

The results of Table \ref{tab:comparison} seem rather unexpected: none of the considered methods is able to predict disease nodes, except $RW^2$ and, with some "caveat" given the adopted evaluation procedure,  DIAMOnD. 
The purpose of next Sections is to interpret more clearly our results.
\subsection{Robustness and sensitivity analysis}
\label{sub:robustness}
\begin{figure}[thpb]
\centering
\includegraphics[width=1.0\linewidth]{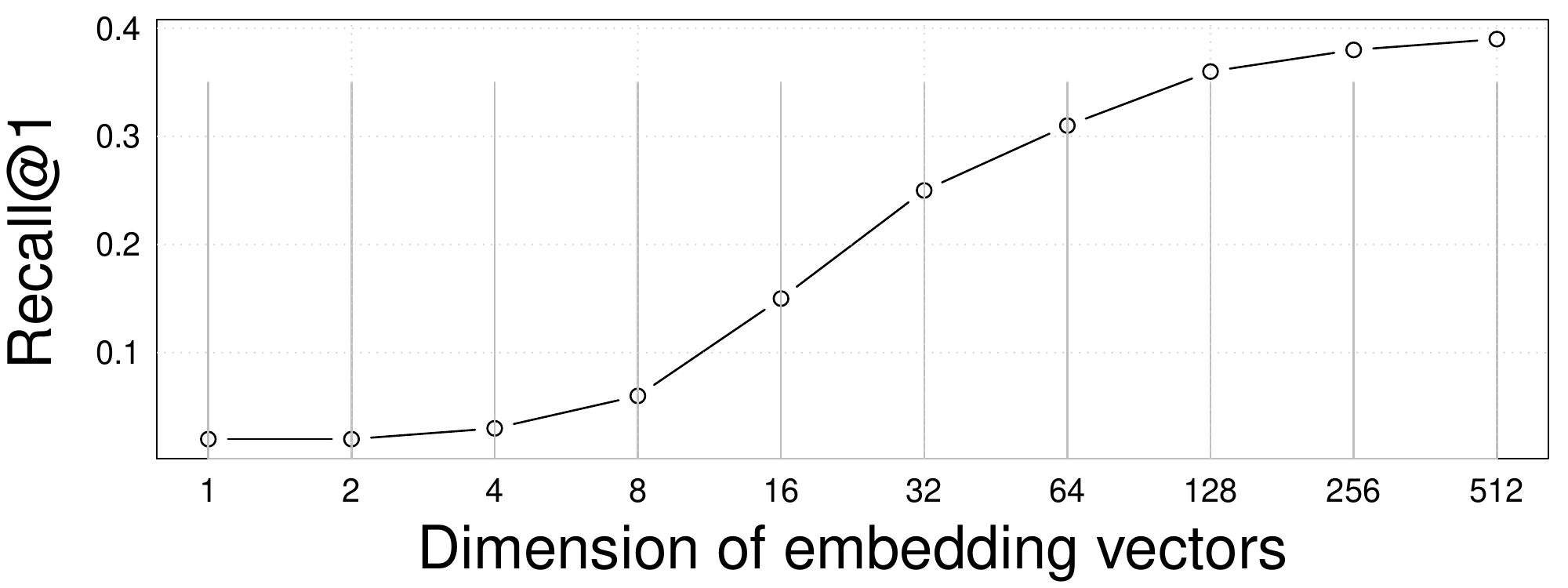}
\caption{Sensitivity analysis: performance as a function of the dimension of embedding vectors.}
\label{fig:emblength}
\end{figure}

\begin{figure}[thpb]
\centering
\includegraphics[width=1.0\linewidth]{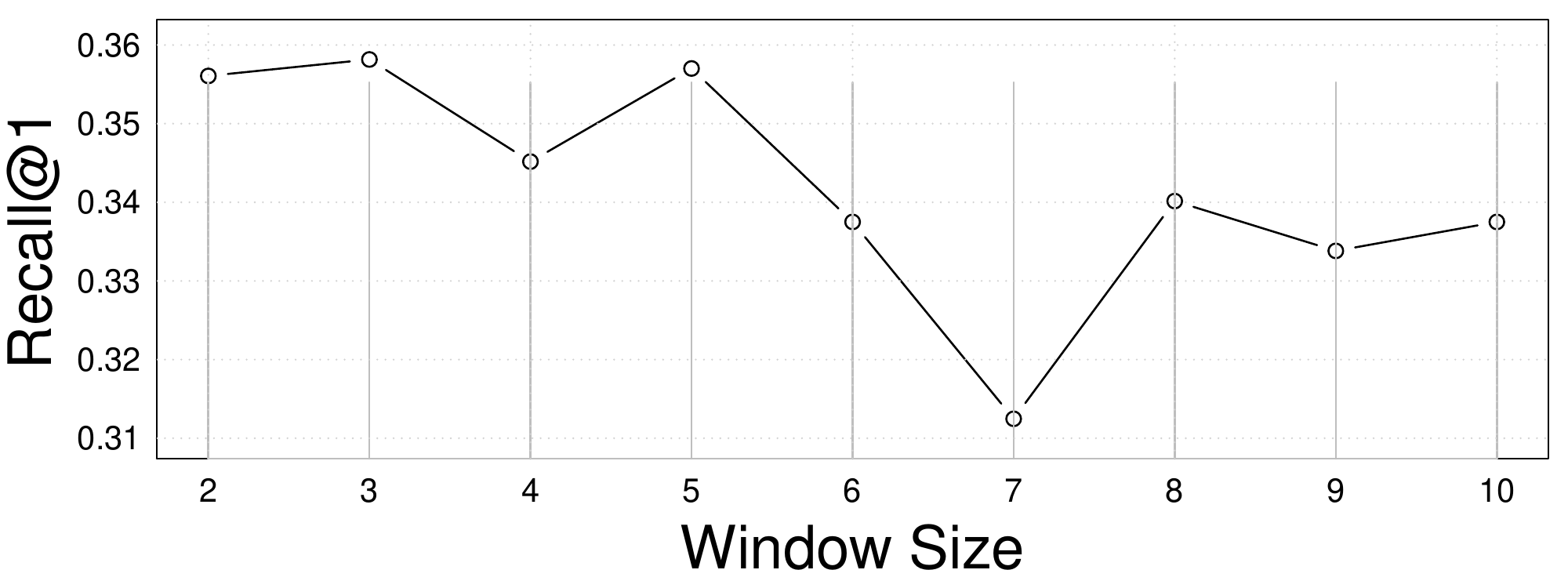}
\caption{Sensitivity analysis: performance as a function of the dimension of Skip-gram context window}
\label{fig:windowtuning}
\end{figure}

First off, we perform a robustness test of $RW^2$, by randomly rewiring edges\footnote{using the method in \url{http://bit.ly/2N0sKHf}}, relabeling node features, and both. The result is shown in Table \ref{tab:robustness}.  The table shows that while relabeling severely affects system's performance, rewiring only results in a 3\% points decrease in performance. Although this difference is statistically significant ($p<0.02$),  it clearly shows that connectivity is a weak feature. This is in line with Table \ref{tab:comparison}, that shows extremely poor performances for methods based only on connectivity patterns (N2V and GCN).  
\begin{table}[H]
\centering
\begin{tabular}{lcc}
\toprule
$RW^2$ (method 2) & R@1 & STD\\
\midrule
Rewiring & 33\% & 0.0086\\
Relabel & 2\% & 0.0005\\
Both & 1\% & 0.0003\\
\bottomrule
\end{tabular}
\caption{Robustness test: rewiring and relabeling the network}
\label{tab:robustness}
\end{table}
Next, we analyze the network sensitivity to parameters.  First, we found that increasing the number of layers of the neural network (step 3 of the pipeline) does not improve results\footnote{Although more experiments with different and more complex learners might be needed, our intuition is that data quality -namely, incompleteness and absence of negative evidence - is too low for deep methods to learn regularities.}. Considering the entire pipeline, only two parameters were found to affect the performance: the dimension of embedding vectors, and the dimension of the window (context) used during the label embedding phase. 
Figure \ref{fig:emblength} shows that a sufficiently high number of dimensions are needed ($>100$), which is in line with other domains in which embeddings have been used. 
Figure \ref{fig:windowtuning} shows that best performances are obtained with smaller left-right contexts (a window size between 1  and 5). This confirms that the diameter of disease modules (remember that disease modules are vaguely defined as an "area" where nodes related to the same disease tend to reside) is relatively small, in line with other studies, for example \cite{agrawal18}, stating that the median distance between components in a module is almost 2.9, and \cite{diseaserelationships} where the diameter of a disease module is estimated to be 1.8 in the average.

\subsection{Discussion}
\label{sec:end}
To summarize, all experiments confirm that connectivity patterns only marginally contribute to identifying gene-disease relationships. This is demonstrated both by the very poor performances of methods, such as N2V and GCN, that only exploit connectivity, and by our robustness tests. Similarly, using only functional features does not produce relevant results, as shown by the baseline method performance.  Instead, $RW^2$ is able to discover \textit{specific combinations of connectivity and functional features that have a higher probability of being found in the vicinity of a node related with a given disease}.
The scarce contribution of connectivity features to the discovery of disease modules contradicts previous results in this domain, but is in agreement with a very recent, although preliminary,  study published by a prestigious institution \cite{agrawal18} were the authors demonstrate that 90\% of disease-related nodes do not correspond to single well-connected components in the human interactome network. Instead, nodes associated with a single disease tend to form many separate connected \textit{components/regions}
in the network. In particular, the authors observe that "current methods disregard loosely connected proteins when making predictions, causing many disease module components
in the  network to remain unnoticed". 
Our study confirms this finding, and demonstrates that $RW^2$  is a better method to capture common features of  such sparse regions: first, the Random Watcher Walker jointly captures connectivity and functional patterns in the vicinity of nodes; second, label embeddings allow to optimize the combination of features types that are more predictive of each disease.
Note that the notion of "vicinity" in embedding methods is more relaxed than "connectivity", since the relative distance between two labels is not fixed, but only constrained by the length of the context window. We also found that performance degrades when the window length exceeds $\pm 5$, which implies that "some" vicinity among nodes related with the same disease does exist.  
\section{Conclusions}
\label{sec:end2}
We have shown that $RW^2$ is able to learn regularities in complex networks characterized by high incompleteness and absence of negative knowledge. 
Our study, together with \cite{agrawal18}, also poses new problems to the Network Medicine field of research. On the one hand, the data quality of the interactome network seems too low for many popular and high performing network methods to produce useful results; on the other hand, the initial hypotheses of Network Medicine are not  denied by our results, since $RW^2$ shows that  some structure in the data indeed exists. Rather, the challenge is to design effective methods to capture at best such sparse regularities (as we have shown, $RW^2$ is a promising solution), and secondly, to enrich or combine in a more effective way different available sources of information. In this work, we use tissues and pathways, since they are a readily available and sufficiently reliable information, however we plan to investigate other sources  in our future work.

\bibliographystyle{ieeetr}
\bibliography{refs}
\end{document}